# BIOLOGICAL ORGANIZATION AND NEGATIVE ENTROPY:

## Based on Schrödinger's reflections[1]

*Francis Bailly*
**Physics, CNRS, Meudon**
bailly@cnrs-bellevue.fr

*Giuseppe Longo*
**LIENS, CNRS – ENS and CREA, Paris**
http://www.di.ens.fr/users/longo

**Abstract**

This paper proposes a systemic perspective for some aspects of both phylogenesis and ontogenesis, in the light of the notion of "biological organization" as negative entropy, following some hints by Schrödinger. To this purpose, we introduce two extra principles to the thermodynamic ones, which are (mathematically) compatible with the traditional principles, but have no meaning in inert matter. A traditional balance equation for metabolism will be then extended to the new notion as specified by these principles. We consider far from equilibrium systems and we focus in particular on the production of global entropy associated to the irreversible character of the processes. A close analysis of this term will be carried on, both in terms of a diffusion equation of biomass over "complexity" and, as a complementary approach and as a tool for specifying a source term, in connection to Schrödinger's method for his equation in Quantum Mechanics. We borrow from this equation just the operatorial approach and, this, in a classical frame, as we use real coefficients instead of complex ones, away thus from the mathematical frame of quantum theories. The first application of our proposal is a simple mathematical reconstruction of Gould's complexity curve of biomass over complexity, as for evolution. We then elaborate, from the existence of different time scales, a partition of ontogenetic time, in reference to entropy and negative entropy variation. On the grounds of this approach, we analyze metabolism and scaling laws. This allows to compare various relevant coefficients appearing in these scaling laws, which seem to fit empirical data. Finally, a tentative and quantitative evaluation of complexity is proposed, also in relation to some empirical data (*caenorhabditis elegans*).

1. **Introduction**

The issue of biological organization, of its emergence, its evolution and of its sustainability has been approached from widely varying perspectives: molecular biology, genetics, open dynamical systems far from equilibrium, etc. However, one of the aspects which remain the most controversial is the thermodynamic one: biological organization may indeed be interpreted in terms of negative entropy, a concept which is absent from physics (where entropy is defined, statistically speaking, using a distribution of probabilities and, macroscopically speaking, according to the direction of heat exchanges). This aspect is the object of debates between several authors among whom we find Schrödinger, Pauling, Brillouin, Atlan, Nicolis and Prigogine. It is to this type of discussion that we would like to return to here, by attempting to introduce different perspectives and methods of approach that are, in our view, closer to the empiricity of life phenomena.

---

[1] Part II of the talk given at the Conference: "**From Type Theory to Morphological Complexity: A Colloquium in Honor of Giuseppe Longo**", Paris, June 28-29, 2007, submitted (see http://www.pps.jussieu.fr/~gc/other/rdp/talks.html).

**1.1 Schrödinger and the entropic heresy**

We will use as starting point Schrödinger's informal and original remarks concerning entropy [Schrödinger, 1944]. Schrödinger's short text is often quoted for its first part, which was quite innovative at the time but is now obsolete. In that part of the text, he proposed to apply the notion of "code-script", even that of program, to chromosomes. Such computational views of the genome have now been made obsolete by many analyses: a synthesis of recent overviews and critiques may be found namely in [Fox Keller, 2003] (see also [Longo, Tendero, 2007]). It must be noted, however, that the notion of program was new at the time, just as was cryptography. Moreover, a Laplacian deterministic viewpoint dominated the period's genomics, and continued to do so for a long time, yet, it had never been explained with such clarity as it had been with Schrödinger. This great physicist, had at least understood the consequences of this application of the discrete symbolism of formal calculus to nature: "It is these chromosomes that contain in some kind of code-script the entire pattern of the individual's future development and of its functioning in the mature state. Every complete set of chromosomes contains the full code. In calling the structure of the chromosome fibers a code-script we mean that the all-penetrating mind, once conceived by Laplace… could tell from their structure whether the egg would develop, under suitable conditions, into a black cock or into a speckled hen… They are the law-code and executive power… they are architect's plan and builder's craft in one" (pp 22-23).

Since the success of the genome project and the decoding of the DNA of several animal species, we have at last arrived to the position of Laplace's God but, unfortunately (?), without the associated predictive power; the least we can say is that we lack the "compiler" and the operating system, even the knowledge of the "executive power". Or maybe is it a case of insufficient knowledge of the global structure within which this discrete sequence operates, a sequence apparently so symbolic and computational, yet embedded in the very complex structure of the cell or even of the organism?

This brings us back to chapter IV of Schrödinger's book where he will "… try to sketch the bearing of the entropy principle on the large-scale behavior of a living organism - forgetting at the moment all that is known about chromosomes, inheritance, and so on…" From this premise, Schrödinger will develop considerations that are as preliminary as audacious and that are based on a view of the organism as a whole. His idea is that what counts for a living organism is its organization and that the problem which poses itself is not only its establishment (the formation of "order based on disorder"), but also its maintenance ("order based on order"). He emphasizes the importance, still unclear today, of the acquisition of organization as negative entropy, including by means of food. This acquisition will participate to the ongoing tension between the increase of entropy, specific to any irreversible thermodynamic process and generating disorder, and the maintenance of order. It is this maintenance of order, its continuous regeneration, actually, that interests us and that we propose to frame '*in abstracto*' by means of mathematical forms as a negative entropy having its own theoretical principles. This will be done independently of any causal analysis which would quite probably require unification with physico-chemical theories; but if we do not have (at least) two theories, with their own conceptual autonomy, there is then nothing to unify.

**1.2 Critiques made to Schrödinger: did you say negative entropy?**

Notions of negative entropy have been introduced in physics on several occasions. Such notions always having been discarded, thermodynamics was able to develop a wide framework without having recourse to anything else than the following general principles (generally in the number of three: the conservation of energy, the non decrease of entropy, the absence of any movement and therefore of any form of energy[2], at the absolute zero). Unification with classical dynamics was then made possible by means of statistics, by the analysis, at the infinite limit, of particle trajectories, hence of geodesics, as optimal trajectories for action (energy × time).

This last remark provides us with the opportunity to highlight that the perspective regarding the inert, in this case just as in any physical theory, is focused on energy and correlated notions: even entropy can be expressed in terms of the degradation of energy (and/or of the dispersion of trajectories). Now, what matters most in the case of living organisms is rather organization: it proposes a crucial observable, from the perspective of the "greater scale" which is the cell or the organism under Schrödinger's examination. Despite the clarity of the choice of scale used by Schrödinger, the reactions to his few pages were as negative as they were uncalled for. Such rejection for many authors would be due to the usual short-circuit: the organization and stability of life phenomena cannot be understood otherwise than in terms of genomic organization and stability. The latter is, of course, of a physico-chemical nature and therefore does not require other concepts than those stemming from physics and chemistry. The very illustrious Linus Pauling is the ring leader for such criticisms: "Schrödinger seems to have asked himself the question: what is the process that leads the production of those well-defined polypeptide chains, with their low entropy?" [Pauling, 1987]. And this is the focus of an article that almost comes to insults ("vague… superficial"): "biological specificity" is, obviously, contained within the "complementary combining regions on the interacting molecules". Now Schrödinger bears no doubts concerning the specificity of the chemistry of living matter, to which he devotes 60 % of his book, but he proposes to *also* look at the level of the whole organism and states so quite clearly: in the above citation he proposes to momentarily forget the chemistry of 'polypeptide chains'.

In [Perutz, 1987], one finds the same critiques and Laplacian certitudes: "the 'one gene, one enzyme hypothesis'… Schrödinger does not seem to have heard of this". This represents a linearity of the causal cascade which leads the author to the conviction that "life can be explained on the basis of the existing laws of physics". Once more, the restriction of a theorization specific to life phenomena is addressed to someone who had already contributed to inventing a *theory that is specific* to microphysics, quantum physics, which is incompatible with the classical and relativistic frameworks. And this is based on an aspect or on a variant of the "central dogma of molecular biology", a very pertinent name for a dogmatic point of view, the "one gene, one enzyme hypothesis". This univocal (oriented) and linear dependency, reminded with great confidence to Schrödinger by Perutz, as well as the dogma, were then proven false by molecular biology itself.

---

[2] At least from a non quantum standpoint.

Given that the same biased interpretation presents itself quite often (when the notion of "organism" arises, "molecule" is the answer); this textbook example must be emphasized, even if this defense of Schrödinger with regard to the "all molecular" perspective partially moves us away from our objectives. It is generally desirable to attentively read what an author writes and to not miss his point, but this is particularly surprising when what we have is a text by a great scientist such as Schrödinger. The physicalistic bias, also having significant impacts for molecular intelligibility, has disregarded any perspective focusing on the organism for at least a century. And *consequently*, this thinking which places the molecule at the center now causes problems: it lacks the ability to grasp the regulating role of the molecule's preferred ecosystem, the cell and the organism - these "reaction norms" of which we hear more and more in molecular biology. In particular, having decoded the genome, we realize that the notion of gene, which has dominated the XXth century, is a concept which is spatio-temporally indeterminate and "causally incomplete": on the one hand, except in particular cases which *do not produce a law* in the physical sense, it is situated within *indeterminate* space and time which extend from DNA to the cytoplasmic activity of RNA, at least in the case of the eukaryote cell; on the other hand, even the primary structure of the proteins generated can be determined by the cytoplasmic context of the RNA transferal (see [Brett et al., 2001], as well as [Fox Keller, 2004] and [Longo, Tendero, 2007] for further references). Besides, the differential method (observation or induction of a mutation, analysis of a consequence, generally teratogenic, at the phenotypical level), specific to molecular biology as well as to physics, requires general principles (in physics, for instance, the geodesic principle, the principles of thermodynamics…), which enable to deduce *direct* and *non-anecdotic* causal relationships between the wild gene and the phenotype.

Schrödinger's viewpoint, which is also ours, is very well explained in [Pichot, 1999; p. 181]: "If we had to summarize the main difference between Morgan's and Schrödinger's genetics, it would be necessary to say that Morgan's genetics attempts to understand how characteristics such as eye color are transmitted from generation to generation, whereas Schrödinger's draws, even within the framework of physics, the problem of the transmission of biological organization. For the one, heredity is considered anecdotally; for the other, it is considered in terms of its foundations. The level of generality is not the same, but this is not what matters most. The main difference is that the one limits itself to seeking to explain the experimental data, whereas the other sets a theoretical problem and looks for its solution."

For all of these reasons, we insist, with many other authors and based on Schrödinger, on the necessity to develop, in parallel to the richness of the analyses of molecular biology, frameworks specific to the activity of organisms. These could propose, even for phenomena that occur within the cell, a *structure of determination* that is more adequate for the physical singularity of life phenomena (see [Bailly, Longo, 2006] for an analysis of the various forms of physical determination and an outline of their relationship to life sciences). In this text, it will not be a question of discussing the current stability of the cell as such, or even that of the organism, as meaningful coherence structures within which to set the causal analyses themselves, if possible, of proteinic cascades. This ambition is specific to the systemic approaches of which the notion of "extended criticality" should be part (see [Bailly, Longo, 2008]); we will take in this regard an even more global view point, by an important change of observables and of parameters with respect to current physical theories. To briefly mention one of these changes, maybe our most important contribution, we will examine the relationship between the evolution of the biomass and that of complexity, as a negative entropy, by taking into account the phylogenetic and ontogenetic levels (we will see that this relationship corresponds to an analysis in terms of diffusion, but within a phase space which is uncommon for physics).

**1.3 The theoretical autonomy of life phenomena and the organization problem**

To conclude this introduction, let's observe that Schrödinger devotes a whole chapter to a question which appears several times in his book: is life based on physical laws? His answer is clear: one must expect the analyses of life phenomena, which "cannot be reduced to the ordinary laws of physics", to have a theoretical autonomy. Moreover, "…classical laws of physics are modified by quantum theory". And in physics, what is sought is the unification of these autonomous theories, located at different scales, and not reduction of the ones to the others; specifically, it is an issue of unifying, and not reducing, their field theories. Likewise, Schrödinger states as an example: "…an engineer who is only familiar with steam engines. After having inspected the construction of an electric motor, he would be willing to admit that the latter functions according to *principles* that he does not yet understand" (and he would do well to not try to reduce them to those of the steam engine : at most, he would correlate or unify them).

In the rest of his text, Schrödinger does not propose any specific principle which is mathematically formalized, but he insists on the necessity of analyses of statistical phenomena, these already being extremely important in physics for understanding thermodynamic entropy. These analyses could help establish correlations with physical theories, among which we have the chemistry of macromolecules. In particular, he believes that it would be necessary to strive towards the unification of two "different mechanisms, which would enable orderly processes, a statistical mechanism producing order based on disorder and the new method, producing order based on order". And there lies, in our view, the complexity of biological phenomena: order as an *organized unity, differentiated and entangled, which creates and maintains itself.* From the seminal works of Prigogine and of several others on thermodynamics far from equilibrium and on self-organization (see [Nicolis, Prigogine, 1977], [Kauffman, 1993]) to the attempt in [Bailly, Longo, 2008]), which analyses structural stability as a coherence structure specific to an "extended critical situation", many try to grasp these aspects of the complexity of living organisms. Concerning the second organizing mechanism ('order from order'), Schrödinger outlines the idea, which we have mentioned, according to which it would use the absorption of negative entropy from the environment, particularly by food. We will not make any assertions concerning the relevance of this idea; however, we will try to propose a partially formalized framework for the notion, strictly correlated in our view, of negative entropy, order, organization, biological complexity.

In particular, we will here take a path which, without exploring the 'causes' – possibly molecular – will attempt to postulate and develop some principles which could help to establish a conceptual framework, possibly mathematical, for the analysis of the role of negative entropy in the play between order and disorder within the living organism, based on the level of the cell (which is, obviously, made up of molecules, in the same way as the classical or relativistic falling bodies are made up of quanta, in their own field: *unification* is indeed progressing today, but it is far from being accomplished[3]).

We will begin by a "principle of establishment/maintenance" of negative entropy, which has no analogy (nor meaning) in current physical theories. We will consider this entropy as a measurement of the organizational complexity of life phenomena; on such bases, we will outline a few mathematical consequences of this identification which will be compared, in a very preliminary manner, to empirical data. Apart from the addition of this new entropic principle, which is specific to life phenomena, and of its consequences on a metabolic balance equation, we will by no means change "the laws" of any physical theory, but we will deeply modify the conceptual space (or the phase spaces) of the considered phenomena and their evolutions. In addition to these equations (or inequalities), the limit case (the value 0 of the components of the "biological type") bases itself on classical physical frameworks, of which these equations are, therefore, nothing but an *extension*.

## 2. Organization as negative entropy: a few principles

From here onwards we will equally use the terms of negative entropy $S^-$ (a negative magnitude) and of complexity K, opposite to $S^-$, (so $K = -cS^-$, where c is a dimensional, positive, constant) which will be a positive magnitude. We will also be led to distinguish processes of complexification in the course of ontogenetic development (internal to the organism and strongly dependent on genetic determinations) from phylogenetic processes of complexification (apparently much more dependent on random phenomena and external conditions).

The initial situation (cell-egg in the first case, isolated bacterium in the other) will be characterized by a very small K (epistemic) complexity (a negative entropy approaching 0, therefore, from a global standpoint, a negligible one). So we then propose as *structural principle* for life phenomena during its organization and the maintenance of its organization the two following inequalities:

$$- K = cS^- \leq 0 \quad and \quad -dK/dt = cdS^-/dt \leq 0 \qquad (1)$$

$S^- < 0$ would correspond to "negentropy" associated to the system's internal organization processes (existence *and* establishment of order, respectively). For purposes of comparison with the physical situation and in order to include life phenomena, we will write the physical entropy corresponding to disorder as $S^+ > 0$. The relevance of this distinction will be clear later on, but let's mention for now that each component is associated to time constants that are sufficiently different to be separated according to the time scale considered (typically, the frequency of metabolic cycles vs. that of cellular reproductions).

---
[3] Let's note, passingly, that even the most elegant of *theoretical reductions*, that of the reduction of thermodynamics to statistical physics, was accomplished when thermodynamics and its principles were already quite solid.

The inequations in (1) thus express, in our view, the principles of the tendency towards organization and to its maintenance within life phenomena, the only context, for us, where non null $K = -cS^-$ would make sense. We will see that the canceling out to 0 of the second equation, in presence of $S^- < 0$, can only concern the accomplished organism resulting from ontogenesis and that it is not achieved in the case of phylogenesis, because, on average, it becomes more and more complex At this stage, we give no (physical nor) proper dimension[4] to $K$, and consider it a "pure number" (see, in particular, $K_f$ below).

To remain closer to empirical reality, in the last part of this text we will consider complexity K to have three main components which can either be of equivalent importance or which can, to the contrary, be clearly dominated by the one or the other according to the situation and we will write:

$$K = \alpha K_c + \beta K_m + \gamma K_f$$

$\alpha$, $\beta$, and $\gamma$ are the respective "weights" of the different types of complexity within the total complexity (we will have $\alpha + \beta + \gamma = 1$). These weights are likely to present temporal variations over the course of an ontogenetic development or of phylogenetic evolution.

$K_c$ ("combinatorial" complexity) corresponds to the possible cellular combinatoric without any other consideration than the differentiations between cellular lineages as structuring element; indeed, inasmuch as cells from a same lineage are interchangeable, it is less their number which is important than the differentiations associated to the apparition of these lineages (although, we will see, their number does intervene). For example, we will consider the analysis of the embryogenesis of *Caenarhobditis Elegans* from this angle (see Appendix 2).

$K_m$ ("morphological" complexity) is associated to the topological forms and structures which arise; it can in principle be mathematically evaluated from the way in which organic structures of a same level of organization present themselves and combine. We will evoke in particular the properties of connexity and more or less fractal structures.

$K_f$ ("functional" complexity) is, for its part, associated to the relationships established and to the fulfilled biological functions; metabolic relations, neuronal relations, interaction networks. In this regard, we will examine in particular the examples from the nervous system on the one hand, and from metabolic networks on the other. In [Edelmann, Tononi, 2000], a measure of biological complexity is proposed, as differentiation of the neural system, by an information theoretic approach, based on Shannon's entropy. This also gives a pure number and it may be seen as a component of our $K_f$.

This tripartition of $K$ ($= -cS^-$), greatly qualitative for the moment, should lead to understand why an increase of K cannot be treated as a decrease of $S^+$, which is physical entropy : K is to be associated to biological organization, particularly to the *alternation of levels of organization*, and to the structurization specific to life phenomena (organites, cells, organs, multicellular biological organism), which is foreign to physical theorizations. As for the instauration of order, critical transitions, studied in physics and acting as starting point for our reflections on "extended criticality" in [Bailly, Longo, 2008], the establishment of coherent structures (percolation, the formation of a crystal, of a snowflake… [Binney et al., 1992; Kauffman, 1993; Jensen, 1998]), corresponds to a decrease of $S^+$, but there is nothing there to allow to speak as such of "different organization levels", nor of the $K_c$, $K_m$, $K_f$ partition introduced above. Once more, the point of this paper is to propose a distinction between the decrease of a specific part of the entropy, due, for example, to a pre-existing physico-chemical potential (molecular interactions, typically, that become actual links because of a decrease of Brownian motion – crystals, snowflakes formation…) and the establishment of biological organization.

---

[4] See however Appendix 1 in case of a specific dimensionality [C] for K and its consequence for other constant ($D_b$, for instance).

As we have already evoked earlier, it is necessary to distinguish the processes of ontogenesis from those of phylogenesis, which, although they may present formal similitudes, are not reducible to ones to the others. Recapitulation theory (ontogenesis would recapitulate phylogenesis) has not really been verified, even if embryos do present, at a given stage of their development, indubitable resemblances in their form and functioning (the morphological "bottleneck"). It indeed appears that the framing of random processes by strong internal (DNA) or external (cell, organism, ecosystem) determinations is very different in each of the two cases.

**3 – Metabolism and negative entropy**

Living matter, beyond its reproductive, generative and plastic capacities, among many others, distinguishes itself by the existence of a metabolism which, on account of various exchanges with its environment and of its internal biochemical reactions, enables it to remain dynamically far from equilibrium and to structurally stabilize the "extended critical situation" which characterizes it. In this paragraph, we attempt to analyze, from a thermodynamic standpoint, the dynamics of this metabolism.

Although the approach proposed here takes on a character which heavily borrows from the concepts of physics, a biological specificity will appear from the moment we take into account the evolutive autonomy of its organization and of the resulting "order", in the schematic and highly abstract form of negative entropy.

So let's consider a system far from equilibrium and note as G its Gibb's free energy. In general, we have G = H –TS, where T is temperature, S is entropy and H = U + PV is the system's enthalpy (U is the internal energy, P and V are respectively pressure and volume). By definition, the R *metabolism*, when it exists (in living organisms for instance), corresponds to the difference between the fluxes of generalized *free energy* entering and exiting through the surface Σ:

$$R = \Sigma[J_G(x) - J_G(x+dx)]$$

So we have:

$$R = - \Sigma dx(div J_G)$$

(in what follows, we will forget the element of volume Σdx, which we consider to be unitary).

Besides, the conservation equation (or the balance equation) is expressed in the general form:

$$-\mathrm{div} J_G = dG/dt + T\sigma$$

where G is an extended expression of Gibb's free energy and where $\sigma$ represents the speed of production of entropy associated to irreversible processes.

So let's return to our distinction which, once more, has no reason to appear in current physical theories, and let's decompose[5] the S entropy in the two different parts which are $S^-$ and $S^+$. With these notations in mind, we then obtain from $G = H - TS$ (see note[6]):

$$R = dH/dt - T(dS^-/dt + dS^+/dt) + T\sigma$$

Moreover, given the relationship between mass and energy, we have $H = aM$ where M is the mass (and a is a coefficient which has the magnitude of a speed squared). So R can be rewritten, by highlighting four contributions to the metabolism: first, the variation of mass, the increase of organization, as a decrease of $S^-$, plus the tendency towards disorder resulting in the increase of $S^+$; then, and crucially, we add the production of entropy $\sigma$ (its speed) due to the *irreversibility of the global process*. We thus have:

$$R = adM/dt - T(dS^-/dt + dS^+/dt) + T\sigma \qquad (2)$$

Equation (2) is the *fundamental equation* which will be the basis of the development of a great part of later discussions. Let's note that the inequalities in (1) are to be read as a "principle" which we propose for a theoretization of life phenomena that is to be *added* to physical (thermodynamic) principles, whereas (2) is a balance equation, based upon classical principles of conservation, yet extended to $S^-$.

Before examining the consequences of this, we will focus on a particularly important term of this equation, $T\sigma$, the inevitable production of global entropy associated to the irreversible character of the processes. More specifically, it is the speed of production multiplied by the temperature ($\sigma$ obviously has the magnitude of an entropy applied to time, so $T\sigma$ is a power). We will take into account the fact that, account taken of all irreversibilities, $T\sigma$ is associated to *all* processes at hand presenting such a trait, *including the variation of negative entropy*, $dS^-$ dt. In a *spirit* that is close to those found in Prigogine's works, whose theorems we will not need to use however, the production of entropy, often considered as a "side effect", in particular near equilibrium, becomes for us one of the main analytical tools.

In the sequel, one of our main observables will be the mass (global, the biomass, or the individual mass). Let's then analyse $T\sigma$ in its relation to the mass. Now, $T\sigma$ is a power and corresponds, approximately, to the product of forces by fluxes (of matter, of energy – chemical, for instance – etc.; a flux is proportional to a force, thus to a mass), and is hence proportional to a mass squared. It can therefore be written, up to a coefficient $\zeta_b$ and a term $T\sigma_0$ as:

$$\mathbf{T\sigma \approx \zeta_b M^2 + T\sigma_0} \qquad (3)$$

$\zeta_b$ is therefore a constant which depends only on the global nature of the living entity under study; it is thus 0 in absence of living matter. We will discuss, for example, the different issues of the biomass and of the mass of an organism. $T\sigma_0$ corresponds to the limit of a purely physical irreversible functioning, that is, one where the *living* mass would be null. This does not apply in biological reality where such mass is at least equal to that of the elementary biological entity which is the isolated bacterium, but it may be relevant for a dead organism, which a decomposing chemical structure.

---

[5] This decomposition is not relevant for purely physical phenomena and remains specific to biological ones. It is one of the reasons for which we consider a "generalized" free energy G.

[6] In a note [Schrödinger, 1944], Schrödinger also proposes to analyze the negative entropy of which he speaks of as (a component of) Gibb's free energy. Of course, the metabolism R is a power.

To use this equation, we will inspire ourselves again from Schrödinger but, this time, regarding his physical methodology and his famous equation from an operational viewpoint.

We will focus on equations (2) and (3), because we will consider them to be specific to life phenomena, as they contain terms that do not have meaning for current physical theories or that, more precisely, *cancel out* when we pass from the description of life to physical phenomena. In (2), it is obviously with regard to $S^-$ and to its variation in relation to time; in (3) the main term belong to our approach to life phenomena and we will give an important role to this equation, a sort of balance between global entropy and biomass. Once more, the inert would be nothing but a limit case, the null value of the observables relative to life. In short, we are "just" proposing an *extension* of current physical theories, as our approach is not incompatible with them, just not reducible. To our physicalist friends in biology, we recall that the quantum field is not only irreducible, but also *incompatible* with the relativistic field – and conversely, so far.

**Intermezzo: Schrödinger's equation and operators (recall)**

One of Schrödinger's great ideas was the introduction of the "wave function" in quantum mechanics. Many aspects characterize the originality of this equation, which has changed the course of microphysics. In our approach, we will highlight here its *operational aspect* that later played a determinant role in quantum physics.

Schrödinger's view, at the time of his equation, centered around the wave function as a description of the quantum state. He came to substitute transformation *operators* to *measured quantities*, specific to the mechanics of classical particles.

To understand, *a posteriori*, this very audacious passage, we consider the following wave function, where **p** is the moment and E is the energy:
$$\Psi(\mathbf{x},t) = \exp(i(\mathbf{p}\mathbf{x} - Et)/h)$$
(it is a solution of Schrödinger's equation for an isolated quantum particle but... this does not matter here).

Since **p** and E appear as coefficients of space **x** and time t, respectively, it is very easy to see that multiplying a spatio-temporal evolution function (this function in particular) by **p** or by E is equivalent to differentiating it with respect to **x** or t, that is $\partial/\partial \mathbf{x}$ and $\partial/\partial t$, respectively (up to a coefficient: i/h).

Thus, to these physical *quantities*, **p** and E, can be associated *differential operators*: the derivative with respect to space and time, respectively, the two parameters of physical evolution. Let's then consider the (classical) law of conservation (Hamilton's equation: total energy is the sum of kinetic energy and of potential energy):
$$E = K_E + P_E$$

More specifically, $E = \mathbf{p}^2/2m + V(\mathbf{x})$, where $V(\mathbf{x})$ is the pertinent potential[7].
Now, if we associate

$$\mathbf{p} \rightarrow -i\hbar\partial/\partial\mathbf{x} \equiv -i\hbar\mathbf{grad}$$
$$E \rightarrow i\hbar\partial/\partial t$$

and to space $\mathbf{x}$ the multiplication by $\mathbf{x}$ or by its functions, such as $V(\mathbf{x})$, we obtain Schrödinger's equation ($\hbar$ is no other than Plank's h divided by $2\pi$ and $\partial^2/\partial\mathbf{x}^2$ is the usual laplacian operator $\Delta$ :

$$i\hbar\partial\psi/\partial t = -(\hbar^2/2m)\partial^2\psi/\partial\mathbf{x}^2 + V(\mathbf{x})\psi$$

($V(\mathbf{x})$ is the potential in x, but its expression is not important for the moment, we will return to this).

The operational association performed may be synthesized, very abstractly, as the application of Schrödinger's operator:

$$\hat{\mathbf{O}}\mathbf{Sch} \equiv \{i\hbar\partial/\partial t = -(\hbar^2/2m)\partial^2/\partial\mathbf{x}^2 + V(\mathbf{x})\}.$$

We propose to follow, *mutatis mutandis,* a similar approach for the very different case we have at hand, relatively to temporal operationality in life phenomena. Let's also observe, following many others, that we can also understand Schrödinger's equation as a *diffusion equation*: it has its "parabolic" form (a quantity diffuses, over time, proportionally to a variation of its gradient in space, plus, if applicable, a source or sink term). It presents however two traits which are essentially different from classical diffusion equations: it operates on the field of complex numbers and not only on the field of real numbers, and the "diffusion coefficient" is itself complex. Let's note that by this approach, Schrödinger invented a phase space which was appropriate to the phenomenal domain which interested him. We will indeed take a similar approach, but basing ourselves however on diffusion laws and then justifying the result by a "Schrödinger-styled" method of operational transformations.

**4 – The "diffusion" of biomass with respect to complexity**

Let's attempt here to explain our strategy, even if it means anticipating certain results and making a few repetitions. Empirical data, to which we will return below, seem to indicate that the qualitatively representative graph of the evolutions of biomass in function of complexity takes on a Gaussian form. Also, we know that there is a relationship between this form and random processes as well as with solutions for diffusion equations. We will therefore write the corresponding equation which we also expect to be interpretable in all its terms from the biological standpoint. Once this stage has been reached, in view of introducing an operational representation, in accordance with what we consider to be an essential property of the temporality of life phenomena, we will look for the metabolism's relevant quantities likely in a self-coherent way to serve as foundation for such operators. To this end, we will follow a method which is similar to that which we have encountered to define $\hat{\mathbf{O}}\mathbf{Sch}$. Our purpose is of using them much more generally later on, by showing that their use may, indeed, characterize a difffusion process in the adequate space, based on the great generality metabolic processes.

---

[7] In the case of the one-dimensional harmonic oscillator, we would have: $E = p^2/2m + kx^2/2$.

Let's now be more specific; we will first attempt to fulfill this program in the case of the evolution of biomass. Why give precedence to the case of biomass? Firstly, it deals with life phenomena as a whole without us needing at this stage to take into account the whole variety of its manifestations; then, and to return to the empirical bases which we mentioned earlier, it so happens that the works by S.J. Gould provide us, as we will see, with a starting point and with a very interesting work direction. We will see that the adequate spaces neither correspond with normal physical space, such as in classical physics, nor with the abstract Hilbert spaces of quantum mechanics, but are rather related to this new basic variable which is complexity K, associated to organization. We may call this complexity phenotypic or, more generally, *epistemic*, in contrapposition to the "objective" complexity of physico-chemical processses (see [Bailly, Longo, 2003] for more on this distinction).

The analytical results will then enable us to return to a "diffusional" character for the basic equation in this new space which is specific to life phenomena. In order to establish these results, we will take inspiration from the aforementioned approach and from works by S.J. Gould, such as presented in [Gould, 1991]. In particular, it will be an issue of modeling two of the main aspects of such work: on the one hand, the idea of random processes of evolution in function of the complexity of life phenomena – and of the quasi-Gaussian aspect taken by the occurrence graph of biomass in function of this complexity (figures 1 and 1') – and on the other hand, that of the existence of what Gould calls a "left wall" which imposes itself upon these processes. This "wall" expresses the impossibility of characterizing life phenomena beyond the elementary level of the bacterium. Random evolution then only takes place "towards the right", meaning in the direction of a higher epistemic complexity than that of the bacterium: in fact, any random walk, bounded on one side, statistically progresses ("diffuses") in the direction opposite to the wall. In other words, the global structure of a diffusion is the result of the local interaction which transitively "inherit" the orientation due to the original symmetry breaking. In our case, where this breaking corresponds to the formation of the first bacteria, there can then be local inversions of complexity, but, on average, it can only increase[8].

FIGURES 1 AND 1' (the "frequency of occurrences" corresponds to our "biomass"):

---

[8] To put it into biological terms, "the spreading of the curve can only be explained by the existence of the left wall and by the multiplication of species; the right part of the distribution is a *consequence* and not a cause of this spreading"… "the notorious progression of life throughout history is therefore a random movement introducing distance between organisms and their tiny ancestors, and not a unidirectional impulse towards a fundamentally advantageous complexity" [Gould, 1991]. Of course, we are only thinking here of biological evolution, while neglecting the last few thousands of years, the short history of humanity's invasion of the planet.

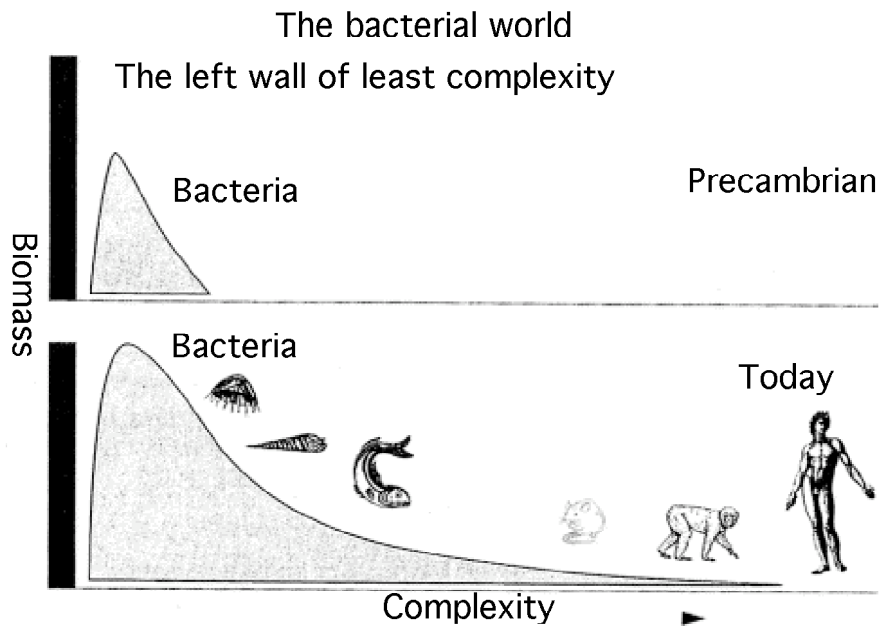

The idea is therefore to define operators (in the field of real numbers, this time) in relationship to this equation between magnitudes which is the metabolism, all the while conserving a character of plausibility between these magnitudes and the operational forms by which they will be put into correspondence. This will enable us to generalize their usage.

**4.1 Dynamics and modelization**

To propose an equation which interpolates, on the basis of general principles, paleontological data, we will use, as observables and reference parameters, the epistemic "density" of the biomass m, physical time t and relative epistemic complexity K: here lies our change of reference space where we will express m in function of t and of K. The isolated bacterium then corresponds to the origin ($K \approx 0$) and the existence of the left wall always imposes $K > 0$, which is consistent with our principle (1). The studied state function will therefore be chosen as the "density" of the biomass relatively to K and will be written m(t, K) of which the integral over all accessible K's will give the temporal evolution of the overall biomass M(t). Time, of course, is an orthogonal dimension relatively to the plane of the figures above: its increase induces a deformation of the curve on this plane, just as in the passage, described by Gould, between the Precambrian time and today.

The dynamics involved and the aspect of the effects which it provokes (Gould's qualitative curve) lead us to propose for m, as a first attempt, a diffusive equation with a source. Indeed, one must take into account an irreversibility with regard to time, an expression stemming from a "random walk" as well as the fact that, by means of growth and genesis, the biomass tends to increase with time. The corresponding "diffusion" equation (which may be interpreted as a balance equation) will thus be written as:

$$\partial m/\partial t = D\partial^2 m/\partial K^2 + Q(t,K) \qquad (4)$$

D represents the "diffusion coefficient" associated to the random evolution process of this biomass density in terms of epistemic complexity and Q is the biomass's source term. The total biomass M(t) at time t will therefore be the integral in dK of m(t, K).

But… how may we justify this equation more specifically and give an expression to Q(t,K)? A Schrödinger type operational approach will enable us to derive this diffusion equation from general considerations made regarding the issue of the production of entropy in metabolic processes and will also enable to propose an expression for the Q(t,K) function.

So let's return to the metabolism equation (2). As we have already recalled, equation (2) in our far from equilibrium frameworks has enabled us to introduce the speed of entropy production $\sigma$, which we have then correlated, by means of equation (3), to the system's energy variation. The latter, let's recall, being proportional to the mass squared, takes the following form in the case of biomass, where M is now the total biomass (as we were saying, T$\sigma$ is a power and the coefficients must, of course, take it into account) :

$$T\sigma = \zeta_b M^2 + T\sigma_0$$

By analogy with what is done in quantum physics regarding energy, that is the association E $\rightarrow$ i$\hbar\partial/\partial$t, it then does not seem artificial to put into relationship the speed of the production of entropy, whch is related to the irreversible character of all processes, with the variation in relationship to time, which is also unidirectional, by means of the partial derivation operator $\partial/\partial$t. Once more, the analysis of the speed of entropy production, far from equilibrium, plays a very important role, from our point of view, one which is quite similar to that of the variations of energy close to equilibrium (see Sect. 3).

So let's set the correspondence - in the manner of Schrödinger, if we may allow ourselves such an abuse of language and… of dimension:

$$T\sigma \rightarrow \rho_b\, \partial/\partial t\,,$$

where $\rho_b$ is a dimensional coefficient (see appendix 1 for the dimensional analysis). <span style="color:red">Similarly as for $\zeta_b$, also $\rho_b$ is different from 0 only in presence of biological activity.</span>

In the same order of ideas, now in analogy to (**p** $\rightarrow$ -i$\hbar\partial/\partial$**x**) in quantum physics, let's then correlate the growing biomass with what may be considered to be its dual or its necessary complement, that is the organization of which it is the locus. Thus, we propose, for K epistemic complexity, and by means of the $\partial/\partial$K differential operator, the following association:

$$M \rightarrow \partial/\partial K \quad \text{(see Appendix 1)}$$

Of course, our parameters and observables as well as the constants (see below) have totally changed: entropy variation, multiplied by temperature, instead of physical energy, mass instead of moment (which is proportional to a mass, though) and, most of all, negative entropy or complexity instead of space. From the formal point of view, and with regard to the physical Hamiltonian, T$\sigma$ then plays the role of energy (it is actually a power) and M plays the role of momentum p (M squared indeed intervenes in T$\sigma$, just as p does in E). Likewise, for Schrödinger, p is associated to the space x, as explained in the Intermezzo, under the form of $\partial/\partial$x, also in relationship to the duality, characteristic of quantum physics, which correlates momentum and position. In our approach, mass is associated to complexity, as the structural organization within the temporal framework where it develops, under the form of $\partial/\partial$K. As we will see below, this component disappears in the equation, exactly when there is only a growth of mass, without any change of organization – in the case of the free proliferation of bacteria, for instance.

In accordance with Schrödinger's approach, the source "potential" $Q(t,K)$ expresses itself, up to a dimensional constant, by the simple multiplication by $T\sigma_0$, which is constant in relation to t and K. This gives us for $Q(t,K)$ a linear expression in m, which we will write as $\alpha_b m$. Intuitively, Q, representing a source term, must be compatible with the tendency towards free proliferation (reproduction) of organisms, which is roughly proportional to the number of existing organisms, therefore, to m (that is, linear in m, see also the following note).

By concluding with the introduction of the "diffusion coefficient", $D_b$, in epistemic complexity, and by posing $\alpha_b = T\sigma_0/\rho_b$ for the source term, we get an operator which takes the form of :

$$\hat{O} \equiv \{ \partial/\partial t = D_b \partial^2/\partial K^2 + \alpha_b \}$$

By using as state function, or "biological evolution function", the density $m(t,K)$ over K, this operator corresponds to the equation:

$$\partial m/\partial t = D_b \partial^2 m/\partial K^2 + \alpha_b m \qquad (5)$$

Of course, $\alpha_b$ makes sense (is non zero) only in presence of non null biological activities ($\rho_b \neq 0$). In the case of the inert, one also has $m(t,K) = 0$. Observe finally that, w.r. to Schrödinger's operator, a crucial difference is given by the coefficients. These happen to be real numbers, not complex ones, as the latter contribute to produce the typical effects of Quantum Mechanics (superposition, among others).

To summarize, in the case of biomass, it was thus possible to associate operators to the relevant magnitudes and to thus obtain a dynamic equation. The recourse to Schrödinger's approach on the one hand justifies, by means of a different method, the same equation obtained as a diffusion, (4); on the other hand, it has enabled us to give an expression to $Q(t, K)$, the source function of the dynamics. One of our preoccupations will now be to examine if, how, and with which results this approach may be applied and generalized to the other cases considered.

## 5. Phylogenetic aspect

We have thus proposed, for the density m of biomass, the evolution equation (5) on the epistemic complexity K, or, more explicitly:

$$\partial m/\partial t = D\partial^2 m/\partial K^2 + am(t,K) \qquad (5)$$

Let's recall that D represents the "diffusion coefficient" associated to the random evolution process of this biomass density over the epistemic complexity and that $am(t,K)$ is the source term of the biomass (D will then have the magnitude of a squared complexity divided by a time value; am is that of the mass density divided by a time value, so a is the reciprocal value of a time value). As intuitively considered above, this would amount to supposing that the proliferation speed ($\partial m/\partial t$) is proportional to the mass[9]. So a solution[10] to (5) would be

---

[9] This hypothesis, to reiterate once again, is perfectly compatible with the analysis of the processes of free proliferation undergone by living matter. That is, with no constraints and without regard to complexity and its variations (so, for D = 0), it leads to an exponential increase in time (for example, in the case of the free reproduction of bacteria).

written as (A is still a dimensional constant, a mass density multiplied by the square root of a time value):

$$m(t,K) = (A/\sqrt{t}) \exp(at)\exp(-K^2/4Dt) \quad (6)$$

To a constant biomass density $m = m_c$, we can thus solve in K to get $K(m_c,t)$, that is :

$$K^2\big|_{m_c} = 4aDt^2 - 2Dt\,\mathrm{Log}\,t + 4Dt(\mathrm{Log}\,A - \mathrm{Log}\,m_c) \quad (7)$$

and for high values for time, the epistemic complexity would increase linearly in function of such time:

$$K(t\to\infty) \sim 2t\sqrt{(aD)} \quad (8)$$

So, for a given biomass density, the epistemic complexity would not stop to increase (and negative entropy would not stop to decrease: $dS^-/dt \sim -2\sqrt{aD}$ ). We could comment this by saying that evolutive processes tend towards a regular increase of the epistemic complexity of a given biomass.

Let's recall in this respect that the "left wall" proposed by Gould for the evolution of the biomass with regard to complexity involves an asymmetry in the whole evolutive process, which is given by the form, asymptomatically exponential in t and Gaussian in K, of our m(t,K) function. As a mater of fact, for purely mathematical reasons, a random walk with a boundary produces an *oriented* diffusion. In our case, this introduces a bias in the variation that is available to selection. So, we obtain an increase, *on average*, in complexity and in mass with the progression of time as well as a (half) bell curve, in what concerns the ratio between the two. This seems to correspond well enough to the empirical evidence and it is contrast to the working hypothesis of the modern synthesis in theory of evolution. According to this hypothesis, the supply of variation, as purely and locally random, is not biased. That is, it was supposed that the variation in a trait is disributed uniformely, in all directions, and without bias around the current mean. As a matter of fact, a simple analysis of the "phylogenetic drift" in terms of random mutations, without principles such as those which we postulate, does not enable to deduce the asymmetry stressed here, following Gould. In fact, the random mutations could induce, at each moment of the evolution and on average, as great an increase as a decrease in epistemic complexity as well as an initially uniform distribution of complexity in relation to mass. Darwinian selection of the incompatible alone would not suffice to explain the overall increase of complexity, because "simpler" may also be compatible with the ecosystem, nor to explain the empirically observed distribution of biomass over complexity. On the other hand, the mathematical justification of asymmetry highlighted by Gould, and that we develop here, accounts for natural complexification: an asymmetry at the origin in the diffusion propagates by local interactions in the phase space. To conclude, it appears to us that these remarks are not necessarily in opposition to (neo-)Darwinian theories, but that they rather insert them within a framework where the structure of evolution is made (more) intelligible by being derived from general principles, among which the (in)equalities in (1) and (2), by very solid methods (diffusion and operator-based approach). In particular, they give a mathematical foundation to the remarks, revisited by Gould and quoted in Sect. 4, remarks which, for many biologists, are at the center of the modern vision of evolution.

---

[10] Other solutions exist, but they do not answer to the constraints that are *a priori* implicit for the object we are examining here.

## 6. Ontogenetic aspect

### 6.1 Three characteristic times and four metabolic regimes

In the case of ontogenesis, the situation is different than that outlined for phylogenesis. Let's start by noting that *embryogenesis,* with the setting of the various functions and a (strong) increase of the complexity of the organism, is completed rather quickly, with a characteristic time which we will call $\tau_K$, to produce an organism which continues to grow without necessarily diversifying further on.

There comes a moment, $\tau_K$, where the negative entropy ($S^- = -cK$) stops decreasing (or where the complexity K stops increasing) and where it stabilizes at the value at which the organization maintains itself (at the cost, of course, of the continuing energetic exchanges with the exterior and of a consumed power P for reaching the final and relatively stable mass). Of course, the setting of the organization is practically over with (end of embryogenesis) a long time before the final mass is attained. Let's call the individual mass W and the characteristic time necessary to reach the adult's mass $\tau_W$ (we will thus have $\tau_W \gg \tau_K$).

In what concerns entropy $S^+$ related to aging, we will propose an exponential increase, with its own characteristic time:

$$dS^+/dt = S^+/\tau_{S+}$$

This increase corresponds, due to the nature of the exponential, to a cumulative effect, with no antagonism (see 6.2.1). The characteristic time $\tau_{S+}$ therefore refers to aging and consequently $\tau_{S+} \gg \tau_W$ because the adult mass is reached far before biologically detectable aging begins.

These three characteristic times ($\tau_{S+} \gg \tau_W \gg \tau_K$) divide the evolution of the organism into the *four distinct periods* below, within which one or another of the relevant aspects is dominant (without excluding the others) : (2.1) establishment of organization (embryogenesis, with a $\tau_K$ characteristic time); (2.2) mass increase ($\tau_W$) ; (2.3) adult life and, finally (2.4), aging ($\tau_{S+}$).

We can therefore distinguish reduced and different forms for the metabolism's equation (2) in function of each of these periods:

(2.1) $R_1 \sim adW/dt - TdS^-/dt + T\sigma_1$  (the effect of $S^+$ remains negligible : embryogenesis)

(2.2) $R_2 \sim adW/dt + T\sigma_2$  (organization $K = -cS^-$ no longer changes, the mass increase continues and the effects of aging remain negligible : childhood/adolescence)

(2.3) $R_3 \sim T\sigma_3$  (now the mass remains more or less constant and all is governed by exchanges with the environment which ensure structural stability: adulthood)

(2.4) $R_4 \sim -TdS^+/dt + T\sigma_4$ (the effect of aging starts to be felt and becomes predominant: old age)

Let's summarize by observing that:
- (2.1) above is the ($dS^+/dt = 0$) case of equation (2) ;
- we go from (2.1) to (2.2), when there is no more increase of organization ($dK/dt = 0$);
- from (2.2) to (2.3), when there is no more increase of mass ($dw/dt = 0$) ;
- from (2.3) to (2.4), when the increase of internal entropy is no longer negligible.

It must also be noted that the (speed of the) production of entropy $\sigma_i$, for $i = 1,...4$, remains present. It could be relevant to consider it as being minimal in $\sigma_3$, at the adult stage – an age of relative "stationarity", but that would lead us to considerations regarding the applicability of Prigogine's "theorem of minimum entropy production" (see [Nicolis, Prigogine, 1977]), which does not affect the work done here.

### 6.1.1 Remarks on aging

Without neglecting the genetic aspects of aging, which molecular theories often associate to the shortening of telomeres, we would like to emphasize the importance of this persistent production of entropy during all the stages of life and, particularly, during the last stages. It is a matter, we reiterate, of the internal entropy $S^+$ which has a physical nature (related to thermodynamic processes and to the exchange of matter and of energy) as well as of the (speed of) entropy production $\sigma_i$ due to *all* irreversible processes, including the $dS^+/dt$ variation of entropy and that specific to life phenomena, the *variation* of complexity $dK/dt = -cdS^-/dt$. Now, in a monocellular organism, for which there are no stages 2.3 or 2.4, given that maturity normally triggers mitosis, the entropy produced is released in the exterior environment and there is practically no reason to speak of aging. On the other hand, in a metazoan, the entropy produced, under all of its forms, is transferred to the environing cells, to the tissue, to the organism. In particular, during the adult stage (2.3) and during aging (2.4) the $\sigma_3$ and, respectively, the $S^+$, $\sigma_4$ components, eminently entropic, dominate. The effect of the accumulation of entropy during life is that which contributes, mathematically, to the exponential increase of $S^+$, with a very large $\tau_{S+}$ (which corresponds to its very tardive sensible manifestation). But entropy implies, in principle, disorganization, including the gradual disorganization of cells, of tissues, of the organism.

But of course, this very general analysis says nothing about *how* this disorganization takes place, nor anything about its specific "timetable". Today, there are at least two competing theories regarding aging (see [Olshansky et al., 2005] for an overview): the first, more classical, based on the cumulative ravages of "oxidative stress", the second, based on the loss of metabolic stability (essentially attributable to [Demetrius, 2004]). These specific analyses account, though differently, for the experimental data and for the observations which are sometimes contradictory. They require, from our standpoint, significant adjustments with regard to our characteristic times, in function of the species and of their ecosystems, but it seems to us that the framework of principles proposed here would be compatible *a priori* with both points of view, yet enriching both, we believe, by their embedding into a more general theoretical frame.

**6.2 Temporal evolution of the metabolism and scaling laws**

In this section, we will compare theoretical observations and empirical data, and this will lead us to a strong hypothesis concerning the correlation between the role of mass, W, and the speed of entropy production, $\sigma$, in the evolution of the metabolism. This hypothesis will be strengthened by a correlation between different magnitudes (coefficients) corresponding to empirical observations.

**6.2.1 Mathematical forms of growth: complexity and mass**

As a premise to Section 6.2.2, the main application of our approach to ontogenesis, we recall that, in order to describe in a mathematically simple way the increase of the individual mass W, in biology, we would usually represent it in the form of the logistic function:

$$dW/dt = (W/\tau_W)(1 - W/W_f)$$

This is the simplest among functions describing an "ago-antagonistic" process (a linear increase which multiplies a decay, the antagonistic factor which limits increases: see the diagram below, in W and t). This factor is normalized by dividing $W_f$, the final mass (asymptotic) reached by the adult organism. In the preceding notation, $\tau_W$ is its characteristic time.

Now, it is legitimate to also give a maximal or final value to the complexity K of a multicellular organism and to also formally describe the evolution of the complexity over the course of ontogenesis, as an ago-antagonistic process, by the logistic function where $\tau_K$ is its characteristic time:

$$dK/dt = (K/\tau_K)(1 - K/K_f)$$

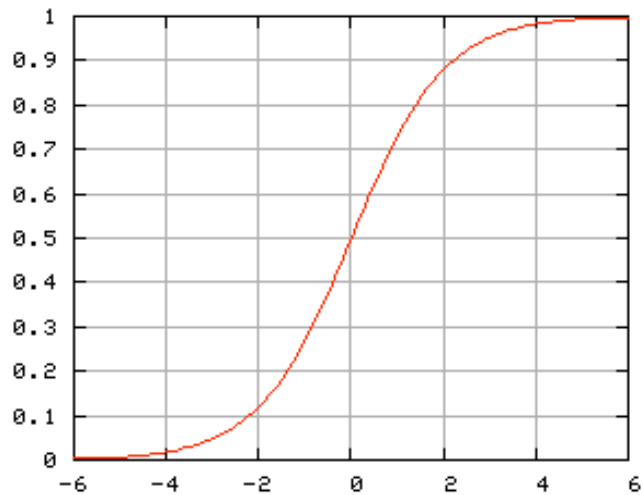

Figure 2.

In other words, we could comment this qualitative diagram, common to dK/dt and to dW/dt, by reminding that, in the case of K, complexity increases over the course of embryogenesis because the structure complexifies and the system becomes increasingly organized. More specifically, after a first phase of simple cellular reproduction, we observe a great increase of organization. This increase slows down, after an inflection point which we have set here at 0, until it then reaches a maximal level of organization, $K_f$, at the end of embryogenesis and of development, here at an approximate "time" 4. We will understand the mathematical form of the increase of mass in a similar way, but with a much longer characteristic time (the two curves only differ by the constant values $\tau_K$, $\tau_W$, $K_f$, $W_f$).

### 6.2.2 The metabolism and scaling laws[11]

The following analysis firstly bases itself on the existence of scaling laws in biology, that is, on the fact that certain magnitudes behave (give or take the coefficients) like powers of the

---

[11] The calculations in this section were established with the collaboration of Boris Saulnier.

adult mass of organisms and sometimes in a very interspecific way [Peters, 1983; Schmidt-Nielsen, 1984]. First, the existence of these scaling laws may have been thrown into doubt, but it happened to be corroborated by a number of observations and reinforced in some ways by the demonstration of allometric laws [Weibel 1991; West et al. 1997; Gayon 2000; Andresen et al. 2002]. Secondly, it is the value of the scaling exponents, generally fractional, which has been the subject of controversies, but a relatively wide consensus, not only based on experimental observations, but also on theoretical constraints, has finally arisen regarding the acceptance of a set of values [Wieser 1984; Denne et al. 1987; Kurz et al. 1998; Gilloly et al. 2001; Andresen et al. 2002; Brown et al. 2002]. It is these values which we will use later on. In particular, for the issue at hand, most of the important characteristic times and biologically important durations (life span, gestation period) or the periods (reciprocal frequency values) that are associated to the biological rhythm scale as the ¼ power of the adult mass, $W_f$. On the basis of this, and for the characteristic times which interest us here, we can write, for $j = S^+, W, K$:

$\tau_j \sim u_j W_f^{1/4}$   (scaling of the times as a ¼ power of the adult mass).

In particular, we have, for the characteristic time of growth:

$\tau_W \sim u_W W_f^{1/4}$.

As for metabolism, the analyses and the observations show that, over *the course of the increase of mass*, it grows linearly with the mass, [Peters, 1983], that is

(R)   $R_1$ et $R_2$ depend *linearily* on the mass W.

On the other hand, the *adult* metabolism itself very generally obeys the scaling law:

$R_3 \sim v_R W_f^{3/4}$.

We will see that the linearity of the dependence of the metabolism over the course of growth and this last equation are correlated: this will only be the limit case of the increase (when the adult mass is reached). In particular, by using $R_3 \sim v_R W_f^{3/4}$ and (2.3), that is, $R_3 \sim T\sigma_3$, we will have:

$T\sigma_3 \sim v_R W_f^{3/4}$.

We will now focus on the presumed linearity of $R_2$, in relationship to W, by comparing it to the expression we get when developing equation (2) at stage (2.2), thanks to the logistic expression which modelizes the increase of mass:

$$dW/dt = (W/\tau_W)(1 - W/W_f)$$

We then get:

(2.2b)   $R_2 \sim adW/dt + T\sigma_2 = (a/\tau_W)W(1 - W/W_f) + T\sigma_2$

Let's firstly note that, for $W = W_f$, we go back, obviously, to the expression for $R_3$. However, we have a problem here: we have just said, see (R) above, that during period (2.2) corresponding to the increase of mass, metabolism $R_2$ linearly depends on the mass, whereas (2.2b) gives a quadratic expression for this dependency.

This apparent contradiction isn't one if the quadratic terms reciprocally cancel each other out, that is, if

(9)   $(a/\tau_W)W^2/W_f \sim T\sigma_2$

From the physical point of view, equation (9) is *dimensionally* plausible, because the speed of entropy production is proportional to the mass squared (see eq. (3)). From a logical standpoint, the inference is correct: if the hypotheses are true and if $(a/\tau_W)W^2/W_f \neq T\sigma_2$ leads to a contradiction, then $(a/\tau_W)W^2/W_f = T\sigma_2$. What appears to validate the hypotheses, in particular (2.2b) which stems from (2), taken together with – or, apparently, … despite – current observations regarding the linearity of $R_1$ and of $R_2$, in relationship to mass, is that our deduction *implies* a relation between the empirically corroborated magnitudes, as we will see in the following section.

Let's note, from this standpoint, that (9) implies, of course, the simplified form of (2.2b):
$$R_2 \sim (a/\tau_W)W.$$
This expression, at the threshold value, when the adult mass $W = W_f$ is reached, that is, when $R_2$ becomes $R_3$, gives us, on the one hand :
$$R_2 = R_3 \sim (a/\tau_W)W_f = (a/u_W) W_f^{3/4} \text{ since } \tau_W \sim u_W W_f^{1/4} .$$
On the other hand, we know that $R_3 \sim v_R W_f^{3/4}$, therefore $v_R W_f^{3/4} \sim (a/u_W) W_f^{3/4}$, and this implies

(10)  $\quad u_W v_R \sim a.$

We will see, in the following section, that $u_W v_R \sim a$ is empirically corroborated at least in the case of the human organism where we have a sufficient amount of data, and this reinforces the given hypothesis.

For the moment, let's use the expressions $R_2 \sim (a/\tau_W)W$, $\tau_W \sim u_W W_f^{1/4}$ and relation (10) to put $R_2$ in the form of

(11)  $\quad R_2 \sim v_R(W/W_f^{1/4})$

At the adult stage, when $W = W_f$, the scaling of $R_3$ in $W_f^{3/4}$ is widely recognized. As promised, in our approach, this scaling becomes a particular case (a particular regime, the adult threshold $W = W_f$) of our more general relation (11). In any event and moreover, we will obtain, in the considered regimes and by using (9) and $\tau_W \sim u_W W_f^{1/4}$, the expression
$$T\sigma_2 \sim v_R(W^2/W_f^{5/4}).$$
This result, at the limit of $W = W_f$, further reinforces our hypotheses and equality (9) which results from it because it gives, by another path, the expression $T\sigma_3 \sim v_R W_f^{3/4}$ obtained above.

In fact, the crucial remark, over the course of this reasoning which is logically and physically plausible, is indeed that the speed of entropy production $T\sigma_2$ (quadratic in W) intervenes in these regimes in such a way as its contribution to the metabolism *compensates* for the "antagonistic" component specific to the mass increase, that is, $-(a/\tau_W)W^2/W_f$ (equation (9) above). It appears that there is something interesting to understand here and which arouses open questions concerning the role of the speed of the increase of entropy, due to all the irreversible processes at play, in the computation of the metabolism in relation to mass. At the $T\sigma_3$ limit, we were saying, and therefore for $W = W_f$, all is in order; meaning, once more, that our general equations, in limit cases, produce widely acknowledged scaling laws.

## 6.3 Comparisons with observations and with biological data

We are now at the stage of verifying that the relation ($a = v_R u_W$), which we have established by using theoretical hypotheses and empirical references, is compatible with the biological data which we may have at hand, the most complete ones seeming to be those relative to the human being. We will proceed in several steps.

Firstly, let's give an explicit expression to the evolution of mass over the course of development; we know that it satisfies the logistic equation and, after integration, we get:

(12)  $\quad W(t) = (W_i W_f)/[W_i + (W_f - W_i) \exp(-t/\tau_W)]$

where $W_i$ and $W_f$ represent the initial and final masses, respectively.

The graph of the growth curve for mass, represented in figure 2, shows an inflection point. We can easily calculate that it is reached at time $t_r$ such that $W(t_r) = W_f/2$.

If the maximal mass of the average adult male is around 70 kg, the usual growth curves show that a child reaches a mass of 35 kg around the age of 12 (= $t_r$). Also, we evaluate the fertilized ovule to have a mass of $W_i \sim 1,4.10^{-3}$ mg (an ovule has on average a diameter of 140μm and a density of approximately 1Kg/dm$^3$). Finally, we use the relationship between $t_r$ and $W_f$ by applying it to equation (12) to get the approximate value of $\tau_W$, that is, $\tau_W \sim 0.5$ year. By then using the scaling formula for $\tau_W$, we finally get $u_W \sim 63$ days/kg$^{1/4}$.

Also, it is possible to evaluate $v_R$. Indeed, the $R_3$ human metabolism is of the order of 100w (100J/sec or 2000 Kcal/day). With $W_f \sim 70$Kg and knowing by the scaling law that $v_R = R_3/W_f^{3/4}$, we /compute $v_R \sim 360$KJ/(day.Kg$^{3/4}$).

Now, if the sought relation (a $\sim u_W v_R$) is verified, we should get a $\sim 22.5$KJ/g. This result is indeed in accordance with the evaluations conducted experimentally by biologists which propose the approximation 20KJ/g < a < 26KJ/g, see [Mitchell, Seymour, 2000; Zelter, 2004].

Of course, the satisfying aspect of this result does not enable in itself to prove the plausibility of the general model which we have just proposed, but gives it nevertheless, besides its relatively simple thermodynamic clarity, a biological plausibility that is not simply abstract. In reality, it would be necessary to complete this sort of result by means of more numerous and general observations and biological experiments. This wold allow to be totally convinced that this approach based on the role of entropy, far from equilibrium, without entering into details of the underlying cycles of reactions, which are of course very important, properly accounts for the metabolic phenomena at the thermodynamic level at which we have considered them for the whole organism or set of organisms.

## 7. The components of complexity

Let's now return to the tripartition of the complexity K introduced in Sect. 2:

$$K = \alpha K_c + \beta K_m + \gamma K_f$$

Our main aim is to propose a "quantitative" approach to "epistemic complexity" of organisms, as we called it. This very arbitrary and sketchy attempt is only justified by that aim, which should turn organization as complexity into a major observable in biology, and a mathematizable one.

Now, the distinction between negative entropy, related to growth as such, and differentiation, related to morphogenesis, have a cellular equivalent. We propose to consider the processes of cellular division and proliferation to be associated to the first aspect and, complementarily, the processes of cellular differentiation for their part to be associated to combinatorial complexity, as well as to consider them to mainly intervene in the context of the establishment of organization under the aspect of $K_c$.

### 7.1 The combinatorial component Kc

We will now roughly simplify the problem by firstly adopting the following combinatorial approach: if N(t) represents the number of cells at time t and if we designate by n(t) the number of differentiated cellular lineages, lineage j (j = 1, …, n) comprising $n_j(t)$ cells ($\sum_j n_j = N$), we will concede that the combinatorial component of the complexity $K_c$ will be given, give or take a dimensional constant, by the logarithm of multinomial $M(t) = N !/\prod_j(n_j !)$, that is:

$$K_c \sim \text{Log}(N !) - \sum_j \text{Log}(n_j !) \qquad (13)$$

Now if we posit that $n_j = N/q_j$, $q_j$ being a bounded integer ($\sum_j(1/q_j) = 1$) and if N is very large compared to 1, we may use Stirling's approximation; (13) can then be simplified and, per cell, we get:

$$K_c/N \sim \sum_j[\mathrm{Log}(q_j)/q_j] \qquad (14)$$

In the Appendix, we illustrate this approach by studying the case of the multicellular organism *Caenorhabditis elegans* of which we precisely know the temporal development, in both terms of number of cells as well as of distinct lineages.

### 7.2 The morphological component $K_m$

In what concerns the morphological complexity $K_m$, we will simply refer to current mathematical analyses which take into account the connexities of organs as well as the existence of critical geometric points (maxima and minima, inflections and curvatures...) characterizing their forms and topologies. For example, and very provisionally, we could evaluate $K_m$ as follows:

$$K_m = n_1 + n_2 + n_3 + n_4$$

where $n_1$ represents the number of critical loci, $n_2$, the number of singular situations (corners, bifurcations, etc), $n_3$, the number of non connate parts of a same organ (for example, the number of separate muscles or bones) and $n_4$, the number of group links in the sense of [Leyton, 2004] which enable to define, for biology, at least closely, the geometric construction of forms. This does not mean, of course, that biology is itself constructed according to these procedures, but that the results of such biological constructions may be described using the method proposed by Leyton… possibly a venue to explore further.

### 7.3 The functional component $K_f$

Finally, the part of $K_f$ complexity which we have called "functional" corresponds to the relations (metabolic, nervous, etc.) which are established between cells and organs in order to ensure the organism's physiology, the integrations and regulations between levels of organization, motor, cognitive and behavioral controls. It is obviously quite difficult to evaluate this contribution but, at the cost of making generalizations which are no doubt too audacious empirically but which may prove to be conceptually useful, we may nevertheless present a few considerations in this respect.

To do so, we will consider that this set of relations and networks can be represented by means of graphs where, for example in the case of the nervous system, the nodes correspond to neurons and the edges to synapses. $K_f$ will then correspond to the (logarithm of) the number of such graphs. So if we designate the number of neurons (approximately $10^{11}$ for the human brain) as m and designate the number of synapses as km (k being between $10^3$ and $10^4$), the theory of Erdös-Renyi graphs shows that there are G graphs such as:

$$G = \binom{\binom{m}{2}}{km} = \binom{\frac{m(m-1)}{2}}{km}$$

with the $\binom{a}{b}$ symbols corresponding to the combination of a objects taken b to b; and we will therefore postulate:

$$K_f \sim \mathrm{Log}\, G$$

If m is very large in comparison to 1, Stirling's successive approximations then give us:

$$K_f \sim km(\log m)$$

And for each neuron, we get: $K_f/m \sim k \log m$

One may notice that the complexity per neuron increases (logarithmically) with the number of neurons once such a number is sufficiently high. This situation may be distinguished from that encountered with the combinatorial complexity per cell which remains – roughly – independent from the number of cells. This effect is of course associated to the global effects that are induced by the functional relations between elements[12].

A more general approach may also be proposed: let $<k>$ be the average number of edges per node and N the number of nodes; the total number of relations will therefore be $<k>N$ and the number of associated permutations is $(<k>N)!$. For a large N the corresponding $K_f$ would therefore be approximately $<k>N \log N$ and, per node (per neuron, for example, or per support within a metabolic network) we get $K_f/N \sim <k> \log N$. with the same qualitative remarks as before regarding the dependency in terms of the average number of edges per node and of the number of nodes. The advantage of this point of view enables to integrate the case of networks which are independent of scale, of which, in general, the probability of edges per node evolves in $k^{-p}$. By taking the normalization factors into account, we get $<k> = \zeta(p-1)/\zeta(p)$ where $\zeta$ represents the Riemann function. A number of studies pertaining to variegated networks show that p is close to 2 (for metabolic networks for example, we have $p \sim 2.2$). For the nervous system, the fact that the average number of synapses is of the order of the thousands, even of the tens of thousands, indicates that in this case p is very close to 2 (give or take $10^{-3}$ or even $10^{-4}$).

## 7.4 Conclusion

We have attempted to define and to analyze what we can designate as the complexity K of a living organism. To do so, we have distinguished between three possible components: a combinatorial component where the important factor is the number of differentiated cellular lineages, a morphological component which takes into account the more or less elaborate form of structures and their connexities and, third, a functional component relative to the relations established by the networks formed by the organism's cells or parts. Depending on the given situation, the dominant terms may vary: for example, in less evolved organisms, the combinatorial aspect, based on the number of cells concerned, may play a major role. Likewise, relatively to the morphological component, the existence of more or less significant symmetries, of more or less numerous connex components, of more or less singular structures (fractal or not) plays an essential role which, in certain cases this may be the main component of (may mathematically dominate) the complexity of the organism. Consider, for example, the variety of organisms involved in the "explosion" of the Burgess fauna. Conversely, in highly evolved organisms, for example those endowed with a sizable and developed nervous system, the relational/functional aspect, logarithmically dependent on the number of concerned cells, seems to clearly dominate.

These different ways in which a same overall complexity K can occur in living phenomena illustrates in our view the genericity of the biological trajectories in contrast with the singular geodesics of physics, inasmuch as this same complexity is, in our view, an essential component of the conceptual space specific to any analysis of life phenomena.

---

[12] This case is the simplest because we have taken into account only combinations of the sets of pairs of interacting neurons. If we had considered the totality (or even an asymptotic significant part of this totality) of the possible sub-sets, we would have obtained $K_f \sim km^2 \log 2$ and a complexity per neuron proportional to m.

# Appendix 1: Some dimensional analyses

Let's recall that the application of the Ô operator leads to the equation of the generic g density function's general form:

$$\partial g/\partial t = D_g \partial^2 g/\partial \kappa^2 + (T\sigma_{0g}/\rho_g)g = D_g \partial^2 g/\partial \kappa^2 + \alpha_g g$$

The dimensional analysis of the various intervening coefficients may present some interest and reveal itself to be enlightening. The dimensions will be denoted in brackets [...], and we will denote, as usual, mass as [M], length as [L], time as [T] (not to be mistaken for temperature which is usually written as [°K]), and, by convention, [C] for complexity. We will then have:

$[\sigma] = [ML^2T^{-3}(°K)^{-1}]$     (power per Kelvin – and per mole -)

$[\rho] = [ML^2T^{-2}]$    (energy)

$[(°K) \times \sigma_0] = [ML^2T^{-3}]$    (power)

$[\alpha] = [(°K) \times \sigma_0 / \rho] = [T^{-1}]$    (reciprocal time value = frequency)

$[D] = [C^2T^{-1}]$    (square of a complexity divided by time).
Let's recall that in the case of thermal or matter diffusion, the diffusion coefficient has a magnitude which is the square of a length divided by time ($[L^2T^{-1}]$; here, it is therefore the epistemic complexity which serves as length, that is, of space. This is in accordance with or main equation (5) and its derivation "a la Schrödinger".

Finally, we have introduced, over the course of these definitions, within the framework of the evaluation of the speed of entropy production, the coefficient $\zeta_b$ ; given the way in which it intervenes (see relation (7) for example), its dimensionality is less "classical"

$[\zeta] = [M^{-1}L^2T^{-3}]$

The other magnitudes which appear in the text are endowed with their usual dimensions (direct: time, mass, numbers, or derived: entropies, energies, densities over the epistemic complexity).

# Appendix 2: The case of *Caenorhabditis elegans.*

The interest of examining the case of *Caenorhabditis elegans* in terms of combinatorial complexity stems from the fact that, as we have already evoked, we have a thorough knowledge of this organism's development, cell by cell and ensuing lineage by lineage. The results we have obtained using the empirical data present some interest and provoke a few questions which may prove to be relevant to other cases.

It is an issue of examining the behavior of $K_c$ over time, which is defined by relation (13) (in this case, there are not always enough cells in each lineage in order to apply approximation (14)). Table 1 presents these results. It is striking to observe that $K_c$ increases very quickly from 0 (a single cell) to 1 and that it stabilizes around this value over the course of its development from the moment where all cellular lineages are represented, as if it was effectively the number of active cellular lineages which would essentially set $K_c$, independently of their number of cells and hence of the size of the organism.

| Time t (mn) | Total number N | AB lineage | MST lineage | C lineage | E lineage | D lineage | P lineage | $K_c$ |
|---|---|---|---|---|---|---|---|---|
| 70 | 6 | 4 | 1 | | | | 1 | 0.57 |
| 100 | 24 | 16 | 2 | 2 | 2 | 1 | 1 | 0.92 |
| 130 | 31 | 16 | 4 | 4 | 4 | 2 | 1 | 1.2 |
| 150 | 81 | 64 | 8 | 4 | 2 | 2 | 1 | 0.71 |
| 170 | 102 | 64 | 16 | 8 | 8 | 4 | 2 | 1.10 |
| 250 | 182 | 128 | 16 | 8 | 8 | 4 | 2 | 0.97 |
| Pre-*lima bean* | 434 | 256 | 64 | 64 | 32 | 16 | 2 | 1.19 |

**Bibliography** (Longo's articles can be downloaded from: http://www.di.ens.fr/users/longo/)

Andresen B., Shiner J.S., Uelinger D.E., "Allometric scaling and maximum efficiency in physiological eigen time", **PNAS**, 99, n°9, p.5822, 2002.

Bailly F., Gaill, F., Mosseri R., "Orgons and Biolons in Theoretical Biology: Phenomenological Analysis and Quantum Analogies", **Acta Biotheoretica**, Vol.41, p.3, 1993.

Bailly F., Longo G. "Objective and Epistemic Complexity in Biology". Invited lecture, Proceedings of the **International Conference on Theoretical Neurobiology**, (N. D. Singh, ed.), pp. 62 – 79, National Brain Research Centre, New Delhi, INDIA , 2003.

Bailly F., Longo G., **Mathématiques et sciences de la nature. La singularité physique du vivant,** Hermann, Paris, 2006 (Introduction in English and in French, downloadable).

Bailly F., Longo G. "Extended Critical Situations", to appear **in J. of Biological Systems**, 2008.

Binney J., Dowrick N.J., Fisher A.J., Newman M.E.J.. **The Theory of Critical Phenomena: An Introduction to the Renormalization Group**. Oxford U. P., 1992.

Brett D., Pospisil H., Valcárcel J., Reich L., Bork P. "Alternative splicing and genome complexity" **Nature Genetics** 30, 2001.

Brown J.H., Gupta, V.K. , Li B-L. , Milne B.T., Restrepo C. West G.B., "The fractal nature of nature: power laws, ecological complexity and biodiversity", **Phil. Trans. R. Soc. Lond.,** B, 357, p.619, 2002.

Chapouthier G., Matras J.-J., "La Néguentropie : un artefact ?", **Fundamenta Scientiæ**, Tome 2, 141-151, 1984.

Demetrius L., "Caloric restriction, metabolic rate and entropy". **J Gerontol Biol Sci** 59: 902–915, 2004.

Denne S.C., Kalhan S.C., "Leucine metabolism in human newborns", **Am. J. Physiol. Endocrinol. Metab.** , 253, E608, 1987.

Edelmann G., Tononi G., **A Universe of Consciousness**, Basic Book, 2000.

Fox Keller E., **The century of the gene**, Harvard U. P., 2000.

Gayon J., "History of the concept of allometry", **American Zoologist**, 40, p.748, 2000

Gillooly J.F., Brown J.H., West G.B., Savage V.M., Charnov E.L., "Effects of size and temperature on metabolic rate", **Science**, 293, p.2248, 2001.

Gould S. J., **Wonderful Life**, Norton & Co., 1989.

Jensen H. J., **Self-Organized Criticality, Emergent Complex Behavior in Physical and Biological Systems**. Cambridge lectures in Physics, 1998.

Kauffman S.A., **Origins of Order: Self Organization and Selection in Evolution**, Oxford U. P., 1993.

Kurz H., Sandau K., **Science**, 281, p.5378, 1998.

Longo G., Tendero P.-E., "The differential method and the causal incompleteness of Programming Theory in Molecular Biology". To appear in **J. of Foundation of Sciences** (extended French version in (P.-A. Miquel ed.) **Evolution des concepts fondateurs de la biologie du XXIe siècle**, DeBoeck, Paris, 2007).

Leyton M. **A Generative Theory of Shape,** Springer, 2004.

Mitchell N.J., Seymour R.S., **Physiological and Biochemical Zoology**, 73 (6), p.829, 2000.

Olshansky S. J., Rattan S. I.S. "At the heart of aging: is it metabolic rate or stability?" **Biogerontology** 6: 291–295, 2005.

Nicolis G., "Dissipative systems", **Rev. Prog. Phys.,** IL, p. 873, 1986.

Nicolis G., Prigogine I., **Self –Organization in Nonequilibrium Systems**, J. Willey, 1977.

Pauling L. "Schrödinger contribution to Chemistry and Biology", *in* **Schrödinger: Centenary Celebration of a Polymath** (Kilmister ed.) Cambridge U. P., 1987.

Perutz M.F. "E. Schrödinger's What is Life ? and molecular biology", *in* **Schrödinger: Centenary Celebration of a Polymath** (Kilmister ed.) Cambridge U. P., 1987.

Peters R.H. , **The ecological implications of body size**, C.U.P. 1983.

Pichot A., **Histoire de la notion de gène,** Flammarion, Paris, 1999.

Schmidt-Nielsen, **Scaling**, C.U.P. 1984.

Schrödinger E., **What Is Life ?** 1944.

Weibel E.R. , "Fractal geometry: a design principle for living organisms", **Am. J. Physiol**, 261, 1991.

West G.B., Brown J.H., Enquist B.J., "A general model for the origin of allometric scaling laws in biology", **Science**, 276, p.122, 1997.

Wieser W. , "A distinction must be made between the ontogeny and the phylogeny of metabolism in order to understand the mass exponent of metabolism", **Resp. Physiol**, 55, p.1, 1984.

Zelter ., "L'homme et son environnement. Aspects bioénergétiques", available at : [www.chups.jussieu.fr/polys/physio/BioenergP1-2004.html](www.chups.jussieu.fr/polys/physio/BioenergP1-2004.html) .